# Physics of stabilization and dynamic control of CW magnetrons for superconducting accelerators

G. Kazakevich#, R.P. Johnson, Muons, Inc, Newport News, VA 23606, USA;
I. Gonin, V. Yakovlev, Fermilab, Batavia, IL 60510, USA;
Ya. Derbenev**, JLab, Newport News, VA 23606, USA.

*Abstract*

Powerful CW magnetrons were suggested for feeding RF cavities of superconducting accelerators because they provide higher efficiency and lower RF power cost than the traditionally used klystrons, IOTs, or solid-state amplifiers. RF amplifiers driven by a master oscillator serve as coherent RF sources. CW magnetrons are regenerative RF self-oscillators with a huge regenerative gain. This causes regenerative instability with notable noise when a magnetron operates as a self-oscillator, i.e., with the anode voltage above the threshold of self-excitation. An injection locking by a small signal was proposed for phase stabilization of magnetrons. In this case, CW magnetrons with injection-locked (coherent) oscillations generate a notable level of noise. This may preclude the use of CW magnetrons in this mode in superconducting RF (SRF) accelerators. This work discusses our analytical results, PIC simulations, and previously obtained analytical and experimental data regarding the operation and control of CW magnetrons. These results led to the development of methods which are most suitable for SRF accelerators.

## 1. Introduction

High-power CW magnetrons, designed and optimized for industrial RF heaters but driven by an injection-locking signal, were suggested in a number of works to power Superconducting RF (SRF) cavities in accelerators due to higher efficiency and lower generated RF power cost than the traditionally used RF amplifiers (klystrons, IOTs, solid-state amplifiers). RF amplifiers driven by a master oscillator serve as coherent, low-noise RF sources. CW magnetrons are regenerative RF self-oscillators with a huge regenerative gain for reliable self-excitation by noise even if the tube is powered by a DC power supply. Very large regenerative gain causes regenerative instability with a notable noise.

Although an approximate theory of magnetron operation, developed by P. Kapitza, was published in 1962-1963 [1], phenomenological models proposed more than 80 years ago are still mostly used. They usually pertain to pulse magnetrons with lower regenerative gain. Typically, the launching of pulse magnetrons produce high harmonics caused by the short rise time of the modulating pulses, unlike CW tubes, starting up from noise and Power-Supply (PS) ripples.

The old models cannot provide the capability to choose CW magnetron operating parameters and designs for application of the tubes in various SRF accelerators.

In [2] were reported first results of the r.m.s. phase deviations of ≈0.8 deg. obtained with a CW microwave oven magnetron with a closed feedback phase loop driving an SRF 2.45 GHz cavity at 2 K.

The measured r.m.s. value includes microphonics phase modulation (≈33 ms period) and modulation with frequency of 60 Hz resulted by the anode power supply ripples and an AC filament power supply.

# e-mail: gkazakevitch@yahoo.com; grigory@muonsinc.com

The general requirement for operation and control of CW magnetrons in SRF accelerators include control in phase and amplitude to meet the accelerator performance providing the r.m.s. deviation in phase and amplitude of ≤0.5deg. and ≤0.5% for Proton/ion accelerators and rings, of ≤0.1deg. and ≤0.05%, in accelerators for Nuclear physics, and ≤0.01deg and ≤0.01% for Light sources [3].

Note that for now the microphonics r.m.s. deviations in phase and amplitude of 0.26 deg. and 0.3%, respectively, at 4 K in the same cavity were demonstrated in [4].

Typically, a CW magnetron is intended to operate as a self-oscillator. Small injection-locking signals, $P_{Lock}$ ≈ -20 dBc or less of the magnetron power $P_{Mag}$ , were suggested to stabilize the tube by a feedback loop. In this case, a CW magnetron may start RF generation from noise or PS ripples, which are uncorrelated with the injection-locking.

Our experiments with microwave oven magnetrons [5] showed that the standard requirement to work above the tube's self-excitation threshold is not a necessary condition for startup and RF generation. The experiments verified startup and stable operation of CW microwave oven tubes fed below the self-excitation threshold at $P_{Lock}$ ≥ -10 dBc.

Here, we summarize the results of our numerical modeling of processes in CW magnetrons, findings from earlier experiments, and PIC simulations indicating Landau damping in a locked (forced-operation) CW tube, as well as innovations developed during our research into the operation and control of microwave oven magnetrons.

A novel approach to magnetron operation and control allows magnetrons to work as coherent RF sources with phase and power control across a wide frequency band, and significantly lower noise spectral density.

Note that the list of references includes the articles in which the noted phenomenon was first discussed.

## 2. Upgraded electrodynamic model of operation of CW magnetrons

We have developed and experimentally tested an analytical electrodynamic model for the startup and operation of CW magnetrons [6]. The model considers a standing wave arising in the interaction space as two slow traveling waves with equal amplitudes rotating around the cathode in opposite directions at the radius $r_S$. The model, based on the charge-drift approximation [1], considers the resonant interaction of the traveling synchronous wave type $exp[-i(n\varphi-\omega t)]$, excited at the frequency $-\Omega=\omega/n$, with the phase velocity equal to $\Omega/k$; $n=N/2$ ($N$ is the number of magnetron cavities, $k=2\pi/\lambda$ is the wave number, $\lambda$ is wavelength of the synchronous wave), with Larmor electrons moving toward the magnetron anode.

Since the wavelength of the traveling wave is much larger than the Larmor radius $r_L$ the interaction of this wave with an electron moving along a Larmor orbit can be treated as the interaction of the wave with a charge located at the center of that orbit. The resonant interaction causes resonant exchange of energy between drifting charges and the rotating synchronous wave in magnetron's interaction space containing a collisionless electron plasma [7]. Thus, we consider influence of self-consistent field interacting with the moving charge, resulting in the Landau damping of the travelling waves [8]. If the phase velocity of the synchronous wave is less than azimuthal velocity of the drifting charge, $v_{az}$, the wave decelerates the charge. This increases the wave energy and the energy stored in the resonant structure leading to phase grouping of charges into "spokes". If the azimuthal velocity of the wave with $\Omega/k$ exceeds the drifting charge azimuthal velocity, the wave accelerates the charge azimuthally and loses energy. This worsens the phase grouping of the drifting charges and may stop magnetron RF generation.

Resonant energy exchange occurs when the azimuthal velocities of the moving charge and the synchronous wave are virtually equal. In this case, the azimuthal electric field of the synchronous wave acts on the moving charge as a constant one, providing the energy exchanges.

In the study of the energy exchange in microwave oven magnetrons were found conditions for phase grouping of charges moving in the interaction space of the tube, necessary for startup and for the most efficient and stable operation at an anode voltage below self-excitation. We upgraded our model to discuss the PIC and analytical modeling, experimental results, and new concepts.

Charge drift in a magnetron is determined by the superposition of the static electric field, described by the static electric potential, and the RF field of the synchronous wave, determined by a scalar potential induced by the magnetron current and the injected resonant RF signal (in a steady state both are in phase). Using the methods described in [9, 6], we obtained equations (1) describing the resonant interaction of the synchronous wave with the charge at the center of the Larmor orbit in the polar frame of the wave, [6]:

$$\begin{cases} \dot{r} = \omega \dfrac{r_S^{\,2}}{r}\varepsilon\phi_1(r)\cos(n\varphi_S) \\ n\dot{\varphi}_S = -\omega \dfrac{r_S^{\,2}}{r}\left(\dfrac{d\phi_0}{dr} + \varepsilon\dfrac{d\phi_1}{dr}\sin(n\varphi_S)\right) \end{cases}, \qquad (1)$$

Here: $\varepsilon = \widetilde{E}_n(r_1)/E_r(r_1) = \widetilde{E}_n(r_1)\cdot r_1 \ln(r_2/r_1)/U$; $\widetilde{E}_n(r_1)$ is the resonant harmonic of the synchronous wave radial electric field at the cathode; $r_S = \sqrt{-ncU/[\omega H \ln(r_2/r_1)]}$, ($H$ <0). $\phi_0(r) = \ln(r/r_1) - (1/2)(r/r_S)^2$, $\phi_1(r) = (1/2n)\left[(r/r_1)^n - (r_1/r)^n\right]$.

Equations (1) show that without the synchronous wave $\varepsilon=0$, and charge drift to the anode is impossible.

The top equation describes the radial velocity of the moving charge. In accordance with this equation, drift of the charge toward the anode is possible in the interval ($\pm\pi/2$) with a period of $2\pi$, i.e., only in "spokes". The charge can enter the spoke through the boundaries located at $n\varphi_S=\pm\pi/2$. The radial drift velocity is proportional to the synchronous-wave radial magnitude, $\varepsilon$. The condition $\varepsilon \geq 1$ does not allow operation of the magnetron.

The second equation describes the azimuthal velocity of the drifting charge in the frame of the synchronous wave. The second term in the parentheses causes the phase grouping of the charge by the resonant RF field via potential $\phi_1$. The first term of this equation describes a radially-dependent azimuthal drift of the charge, resulting from the rotating frame, with azimuthal angular velocity $-\Omega=\omega/n$. This term at $r > r_S$ and low $\varepsilon$ causes movement of the charge from the phase interval of "spokes".

Equations (1) were numerically simulated for a typical model of a commercial magnetron with $N = 8$, $r_1 = 5$ mm, $r_2/r_1 = 1.5$, $r_S/r_1 = 1.2$ [6]. Charge trajectories at $r \geq r_1 + r_L$ were obtained for various magnitudes $\varepsilon$ of the RF field in the synchronous wave and during the time interval of the drift $2 \leq \tau \leq 10$ cyclotron periods $1/\omega$, allowing coherent increment of the synchronous wave.

The azimuthal boundaries of the charge drifting in a spoke at various $\varepsilon$ are plotted in Fig. 1(a) assuming a uniform phase distribution of the emitted electrons. The phase interval at $(r_1+r_L)$ normalized by $\pi$ shows part of the emitted electrons (in the phase interval admitted for a spoke) contributing to the synchronous wave [6].

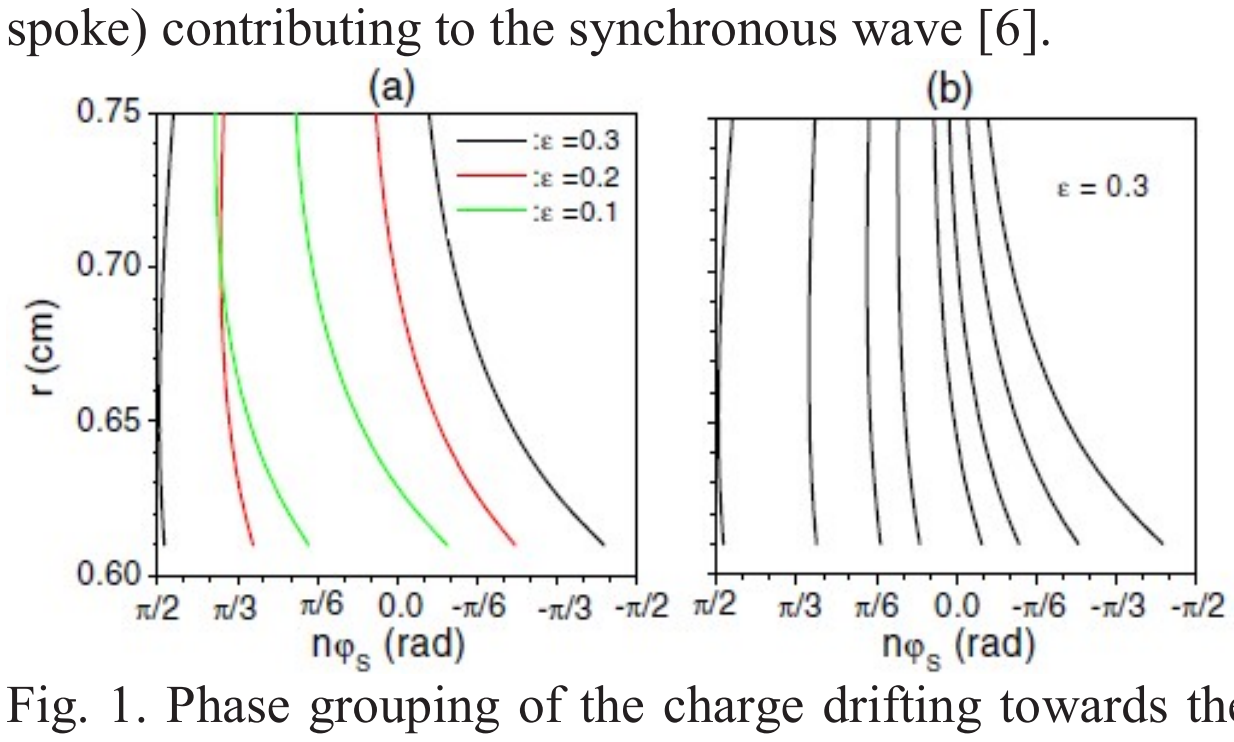

Fig. 1. Phase grouping of the charge drifting towards the magnetron anode in the considered magnetron model. Graph (a) shows the phase boundaries of trajectories of the charges contributing to the synchronous wave in dependence on $\varepsilon$. Graph (b) shows trajectories of the charges in a "spoke" at $\varepsilon=0.3$, i.e., in fact a phase grouping of charges.

At ε ≥0.3 moving charges at all radii within the space of interaction contribute to the synchronous wave [6].

Phase grouping of the drifting charges caused by resonant interaction with the rotating synchronous wave violates the phase uniformity of the initially emitted charges [6]. This phenomenon may be explained as the appearance of a degree of freedom of the drifting charges at the resonant interaction [10]. Taking this into account and differentiating the bottom equation of system (1), one can obtain an "azimuthal acceleration" of the drifting charges shaping the "spokes" vs. $n\varphi_S$ at the radius of rotation of synchronous wave $r_S$ for various $\varepsilon$ as it is shown in Fig. 2 [10].

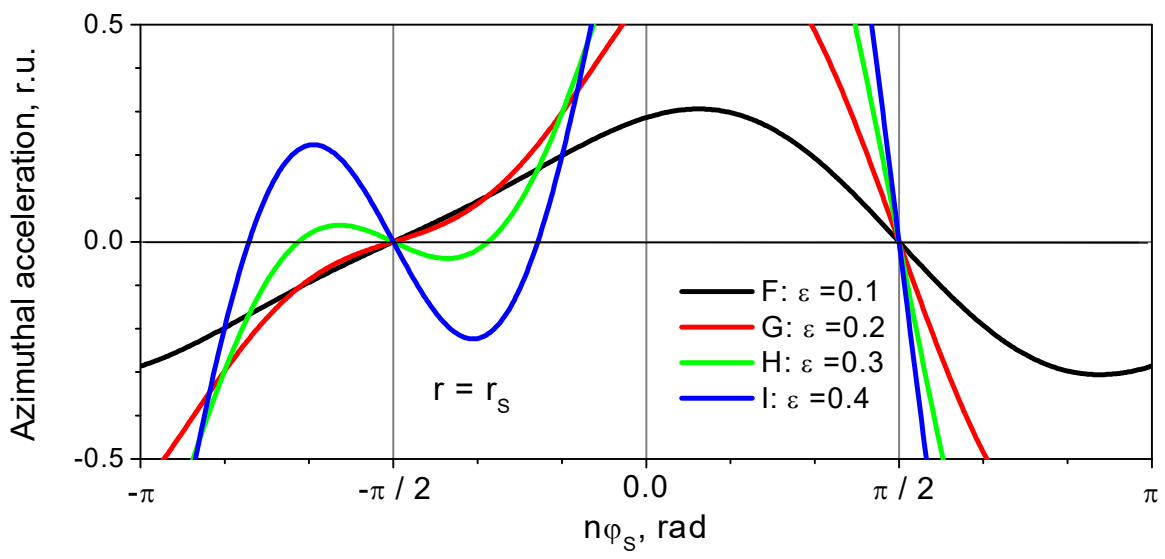

Fig. 2. An "azimuthal acceleration" shaping "spokes" of the drifting charges vs. $n\varphi_S$ at the radius of rotation of synchronous wave equal $r_S$ for various $\varepsilon$.

This plot can be interpreted as the impact of a gradient of the grouping potential $\psi(r, n\varphi_S, \varepsilon)$ of the synchronous wave, with respective potential wells at the radius of rotation of the synchronous wave $r_S$, Fig. 3 [10].

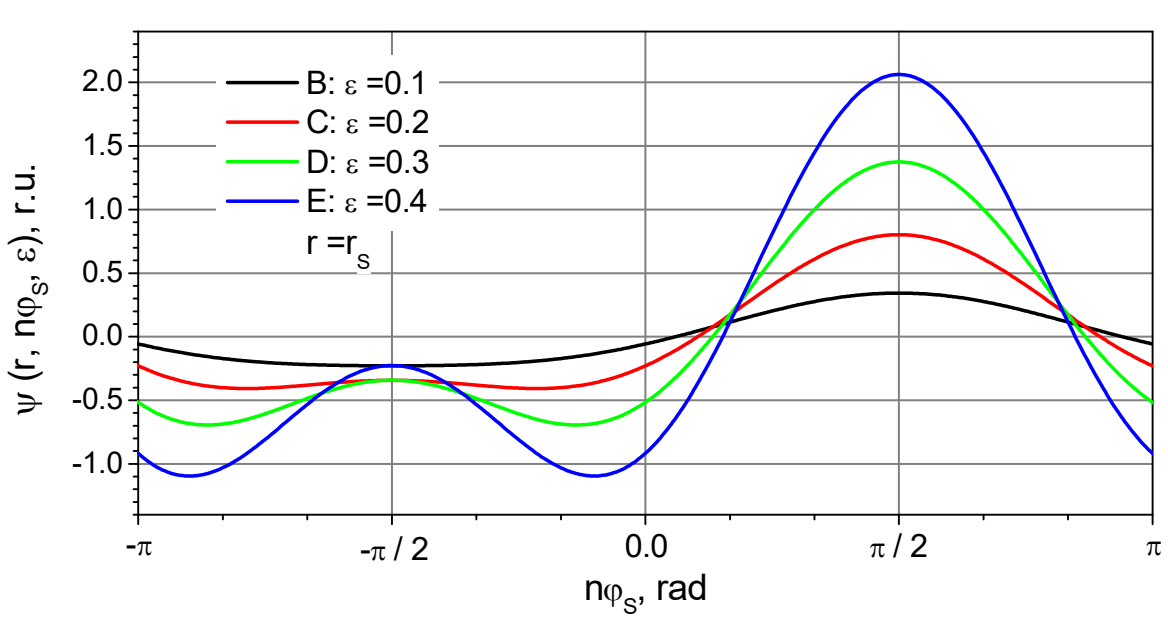

Fig. 3. A gradient of the grouping potential $\psi(r, n\varphi_S, \varepsilon)$ at the radius of rotation of synchronous wave $r_S$, vs. $\varepsilon$ .

This graph shows losses of charges from the "spoke" at $\varepsilon$<0.3 because of migration of part of the charges from the phase interval ($\pm\pi/2$) allowed for a "spoke". This causes electron back-streaming in traditionally used regimes of CW magnetron operation. The emigrated electrons absorb RF energy from the synchronous wave, heating the cathode surface. Note that, in the developed electrodynamic model, at $\varepsilon \geq 0.3$ (at a forcing signal of ≈ -10 dBc or more), the potential barriers prevent migration of charges from spokes for all radii of moving charges, as presented in Fig. 4 [10]. This reduces overheating of the cathode surface by the quite strong phase grouping in magnetrons operating with a quite large forcing signal [10].

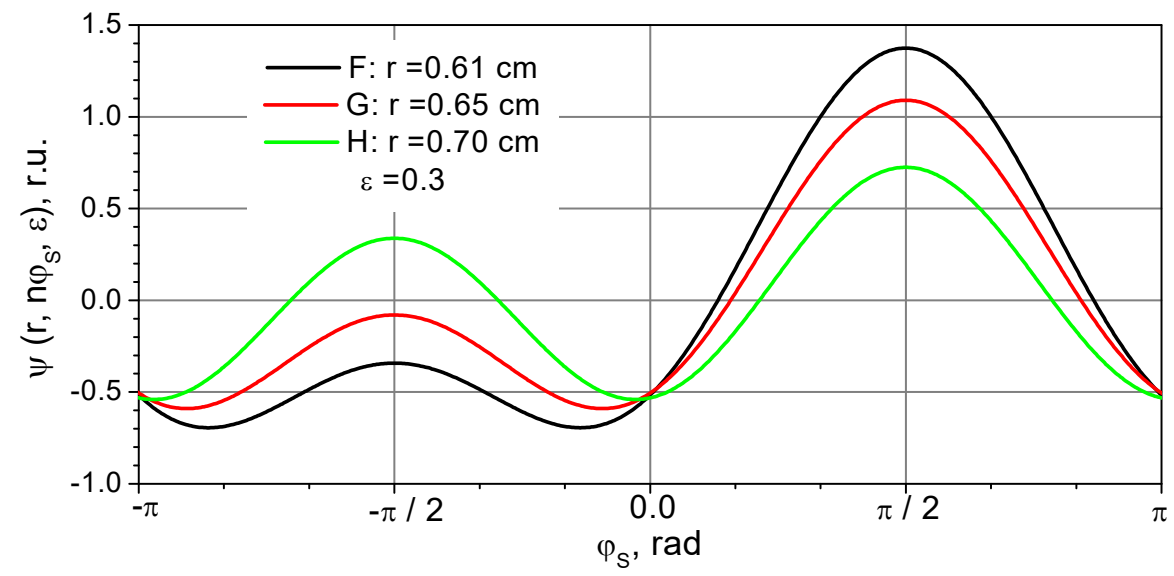

Fig. 4. Potential barriers preventing electron back stream at various radii of the drifting charges in a magnetron model at a large forcing signal $\varepsilon$ =0.3.

Thus, the emergence of an additional degree of freedom associated with phase grouping of the moving charges into "spokes" rotating in the magnetron interaction space due to resonant interaction indicates a noticeable reduction in the back-stream of electrons that overheats the emitting surface when the tube operates with a sufficiently large forcing signal in CW or pulse modes. The PIC simulation [11] performed for a microwave oven magnetron verified an approximately twofold decrease in the power of back-streaming electrons overheating the cathode [10].

The electrodynamic model substantiates previously unknown, but most suitable for various SRF accelerators, methods for the operation and control of CW magnetrons with a relatively large (≈ -10 dBc) injected locking (forcing) signal, allowing a wide range (≈10 dB) of power control by varying the magnetron current [5]. The experiments showed that the magnetron power depends linearly on the tube current with good accuracy; the efficiency of the magnetron with this method of wide-range power control is the highest [ibid.].

The measured V–I characteristic of the CW, 2.45 GHz, 1.2 kW magnetron type 2M137-IL (from Richardson Electronics) demonstrates a current (RF power) control range of more than a factor of 10 at a forcing signal of 100 W [6].

## 3. Phase grouping mechanism in magnetrons

CW magnetrons fed at a voltage notably below the self-oscillation threshold cannot be launched by noise or PS ripples. However, spontaneous RF oscillations exist in the interaction space, as shown in the following chapter. The source of these spontaneous oscillations can be non-stationary oscillations of alternating quasi-static fields in the electron cloud of space charge around the cathode [9]. In addition, the cyclotron oscillations of Larmor electrons with power $P$ in free space [12], per Larmor revolution, may be considered as spontaneous oscillations using, in the first approximation, Eq. (2) for non-relativistic electrons.

$$P = \frac{2e^4 H^2 v^2}{3m^2 c^5} \tag{2}$$

Here: $H$ is the magnetron magnetic field, $v$ is the velocity of a Larmor electron, and $m$ is the electron mass.

For a 2.45 GHz microwave oven magnetron with $H$ ≈880 Oe at an anode voltage ≈3.7 kV, $v \approx 0.097c$, the value of $P$ is $\approx 1.14 \cdot 10^{-18}$ W [13]. For a magnetron with an emitted current of ~1 A, one can expect phase-locked spontaneous-oscillation power $P_S$ of up to ~1 W. Note that Eq. (2) gives the radiation power of a Larmor electron in the far zone. For a magnetron, this is only a very rough approximation. Measurements of the spontaneous oscillation power are considered in the following chapter.

The phase grouping can be characterized by the grouping coefficient $C_{PG}$, numerically equal to the ratio of the azimuthal phase size of the "spoke" with emitted electrons to the phase size of the "spoke" vertex near the anode. The charge trajectories in the "spokes" in a CW tube with 10 cavities obtained by PIC simulation [13], show $C_{PG} \approx 3.8$.

Improvement of the phase grouping increases the fields of the synchronous wave, which in turn further improves the phase grouping, and so on. This means that phase grouping in the magnetron interaction space is a fast, non-stationary, avalanche-like process.

The build-up of a synchronous wave in such a process can be sought as an exponential dependence on $C_{PG}$. Since the azimuthal size of the phase-grouped charges near the cavity slit is much less than the wavelength of the synchronous wave, phase grouping leads to coherent amplification of the synchronous wave with gain $G_C$.

The synchronous-wave magnitude is self-amplified by the coherent gain $G_C$. This strengthens the phase grouping of charges in "spokes" and excites the resonant system at the frequency $-\omega = \Omega \cdot n$, improving phase locking of the spontaneous oscillations. The coherent oscillations enter the magnetron resonant system with regenerative gain $G_R$. The total gain for the oscillations is $\sim G_C \cdot G_R$ [14].

Assuming that the coherent oscillations power has a quadratic dependence on the magnitude of radiating charges and taking into account the exponential dependence of the avalanche-like process of the phase grouping, one estimates the $G_C$ magnitude from the following equation [14]:

$$d(\lg G_C) \sim 2 \cdot I \cdot C_{PG} \cdot dI \qquad (3)$$

Here: $I$ is the magnetron anode current, $dI = I \cdot C_{PG}$ is the phase-grouped current inducing the synchronous wave.

The product $I \cdot C_{PG}$ is dimensionless.

Thus, the phase-grouped Larmor electrons produce the coherent, self-amplified, injection-locked radiation (until saturation is reached) [14] with the maximum $G_{Cmax}$ value:

$$G_{Cmax} \sim \exp[2 \cdot I^2 (C_{PG})^2] \qquad (4)$$

Estimates of the coherent gain $G_C$ vs. $C_{PG}$ in the interaction space of a microwave oven magnetron for $I$ =0.4 A are shown in Table 1, [14].

Table 1. Estimates of the coherent gain $G_C$ vs. $C_{PG}$.

| $G_{PG}$ | 2 | 3 | 4 | 5 |
|---|---|---|---|---|
| $G_{Cmax}$ | 3.6 | 17.8 | 170 | $3 \cdot 10^3$ |

The $C_{PGmax}$ value is determined by PIC simulation as a maximum of the phase grouping; for estimates usage of averaged value of $G_C$ as $G_{Cmax}/2$ looks acceptable.

Thus, phase grouping can compensate for the reduction in regenerative gain resulted from reduced anode voltage by increasing the coherent gain.

## 4. On stable operation of a CW magnetron below the self-excitation threshold voltage

The carrier-frequency offset of a CW magnetron of type 2M137-IL, stably operating below the self-excitation threshold at an output power of 100 W, was measured as a function of $P_{Lock}$ in experiments at RBW = 100 kHz (the anode voltage was ≈150 V below the self-excitation threshold of 4.04 kV), with Ref. = -45 dBc [5], for $P_{Lock}$ = 30, 50, and 100 W.

In the measurements, peaks from the locking (forcing) signal are present at the output in all traces, since the signal is applied to the tube output. However, some peaks may not be caused by magnetron startup. In this case, the oscillation offset can be obtained by eliminating the injection-locking signal from the output power peaks using adjacent averaging, which reduces sharp peaks while preserving the smooth spectra shown in Fig. 5 [14].

In the following graph, the trace with $P_{Lock}$ = 30 W shows broadband RF oscillations in the interaction space with a bandwidth of ≈7 MHz (determined by the tube resonant-system bandwidth) at the -3 dB level, amplified by the resonant system up to an output power of ≈1 W.

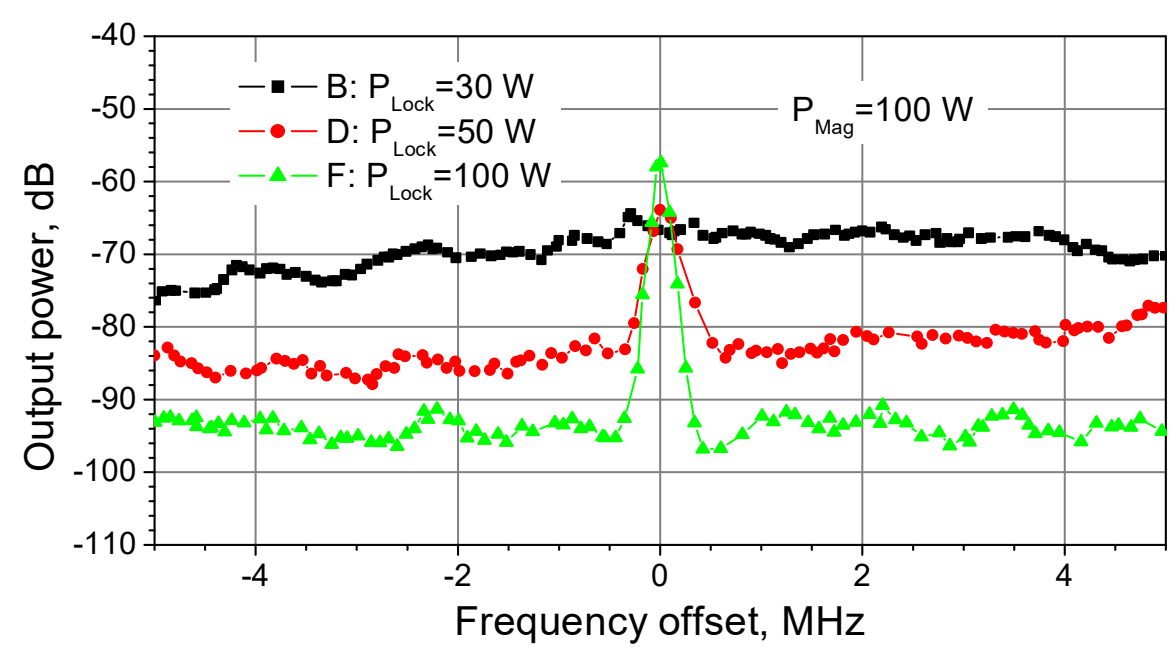

Fig. 5. Adjacent averaged offsets of the carrier frequency of the 2M137-IL magnetron at the magnetron output power of 100 W, Ref. -45 dB, vs. power of the injection-locking (forcing) signal, $P_{Lock}$ [14].

One can infer that the black trace shows amplified incoherent spontaneous RF oscillations close to the launch point, i.e., in effect, the quasi-continuous noise of the unlaunched tube. The other traces show significantly reduced incoherent spontaneous oscillations due to their conversion into coherent oscillations when the magnetron is launched [14]. At $P_{Lock}$ = 50 W, the spontaneous oscillations become partially coherent (partially phase-locked) owing to the stronger azimuthal electric field of the synchronous wave, which causes some phase grouping. The residual spontaneous-oscillation power characterizes the loss of coherence vs. $P_{Lock}$ during phase grouping. A further increase of the locking (forcing) signal to 100 W additionally reduces the power of the spontaneous oscillations by converting them mainly into coherent oscillations, as shown in Fig. 6 [14], where the spontaneous noise is reduced by ≈30 dB.

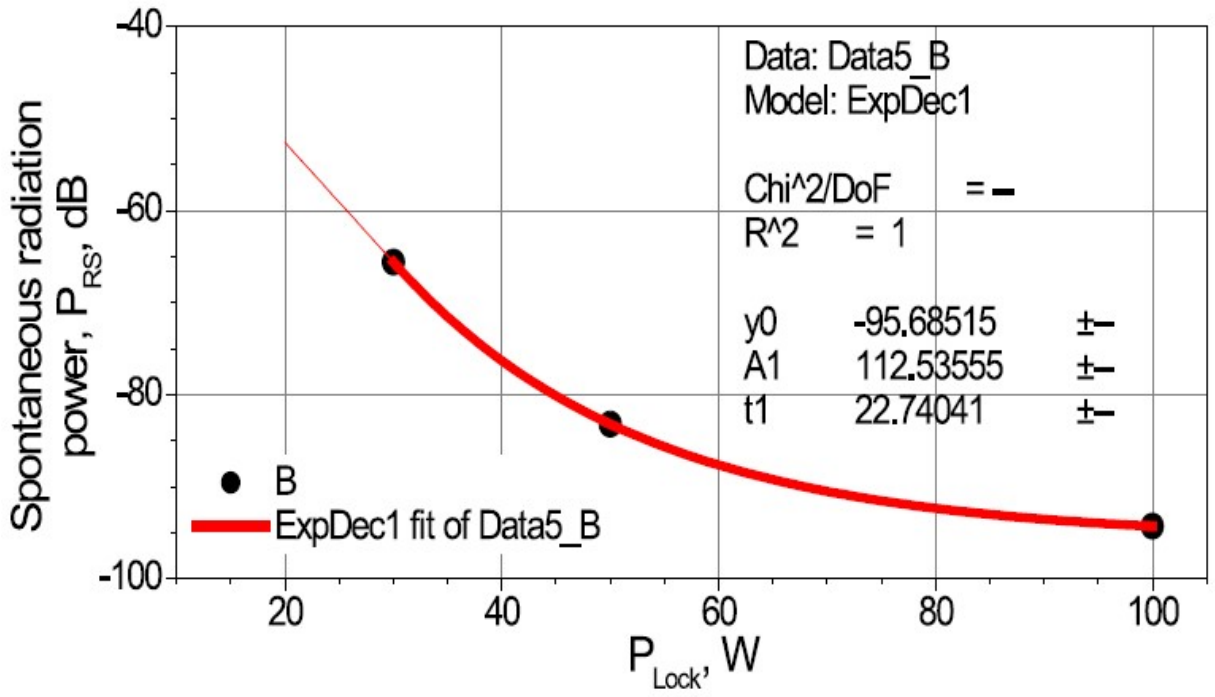

Fig. 6. Fitting the power of the smoothed spontaneous oscillation, shown by dots, by the exponential decay vs. power of the injection-locking (forcing) signal $P_{Lock}$, bold line; thin line shows extrapolation [14].

A noticeable frequency shift of the incoherent spontaneous oscillations at $P_{Lock}$ = 50 W may indicate frequency pushing caused by the anode current loading the resonant system of the launched magnetron.

Thus, in CW 10-cavity microwave oven magnetrons, complete phase locking can be realized at a locking (forcing) signal power of ≥ -10 dBc.

Further, using the electrodynamic model [6], techniques for forced RF generation of CW magnetrons that may eliminate startup driven by regenerative noise and PS ripples were developed and described. These techniques were verified experimentally with CW microwave-oven magnetrons, as shown below.

Particle-in-cell (PIC) modelling [10] of a CW magnetron (frequency 2.45 GHz, power 0.86 kW, nominal anode voltage 4.0 kV), at a magnetic field of 0.18 T and a synchronizing (driving) locking signal level of -8 dBc performed first with high temporal resolution, demonstrates RF oscillations at the magnetron output (top and bottom graphs).

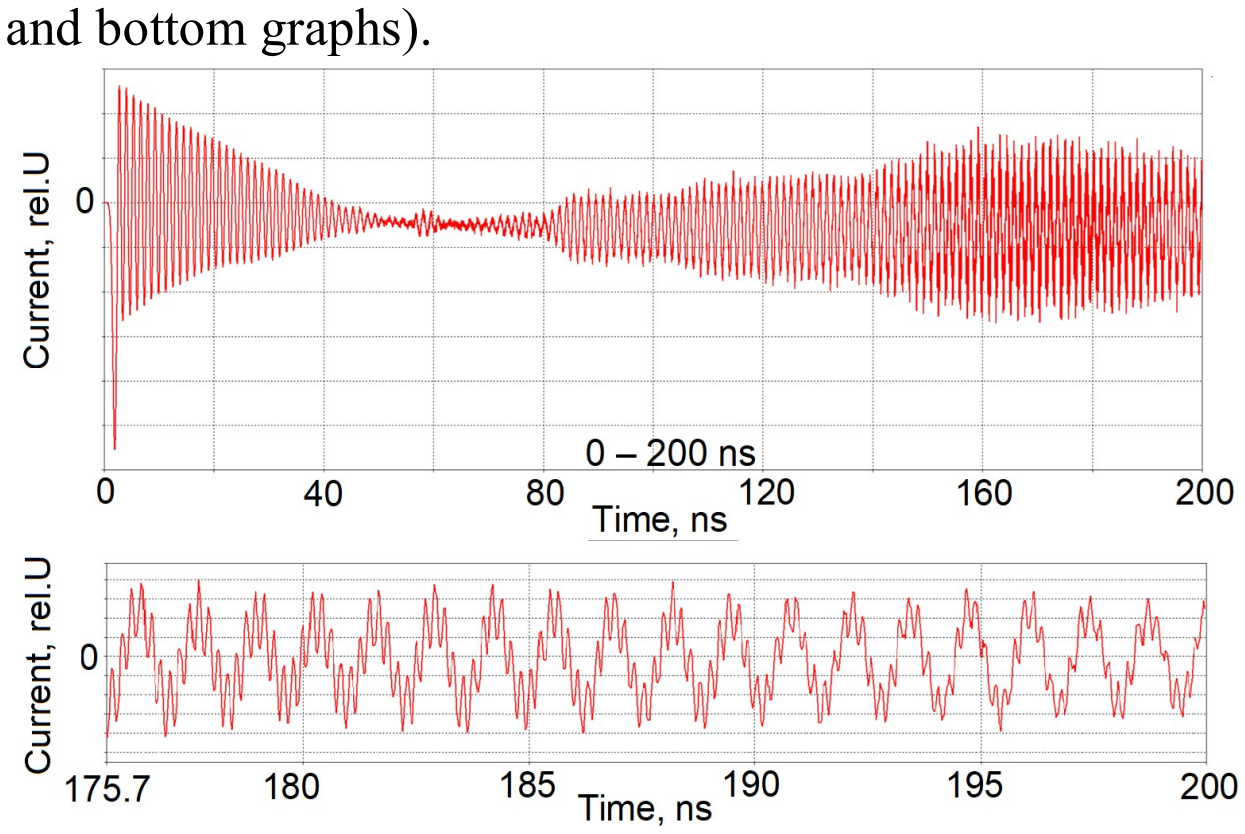

Fig. 7. Upper graph. PIC-simulated oscillations indicating damping of the synchronizing wave and slow build-up at the magnetron output at the forcing signal of -8 dBc and the tube anode voltage of 4.0 kV. Plot shows a transient (quite long) process of injection-locking.

Lower graph, with a delay of ≈176 ns, (this is ~400 periods of the magnetron generations) shows the oscillation caused by currents in "spokes" (at $-\Omega = \omega/n$) and the emergence of locked (forced) oscillation at frequency $\omega$.

The top graph shows the very rapid excitation of a synchronous wave (occupying approximately one rotation of the "spokes" formed as a result of phase grouping under the influence of the forcing signal), as it was noted in Chapter 3.

Clearly seen is the simultaneous occurrence of the processes:

a) Damping (lasting approximately 50 ns) of the synchronous wave that facilitates resonant collisionless interaction with the electron plasma formed by Larmor electrons (which are characterized by a spread in cyclotron frequencies within the tube's interaction space due to magnetic field non-uniformity); this process corresponds to wave damping in plasma (Landau damping) [7].

b) Growth of oscillations caused by the gradual accumulation of RF energy in the resonant system of the magnetron, which in turn affects the amplitude of the synchronous wave.

Some irregularity in the formation of the synchronous wave in the 140–180 ns interval may be caused by quasi-continuous noise from the tube operating above the self-excitation threshold voltage.

The criteria $\Omega/k < v_{az}$ or $\Omega/k > v_{az}$ for the synchronous wave indicate increase or reduce of the synchronous wave propagating in a collisionless electron plasma of magnetrons due to the Landau damping [7]. These criteria characterize the influence of the self-consistent field on the resonant interaction when the synchronous wave propagates in the electron plasma within magnetron's interaction space.

Quite long time interval (~30 ns) when the magnetron oscillations are absent may be caused by significant dumping of the synchronous wave; the interval includes a filling time of the magnetron resonant structure ~ $Q/\pi f$.

An increase in the forcing signal, as shown in Fig. 8, reduces duration of the damping process of the synchronizing wave, and the buildup time of the injection-locked oscillations.

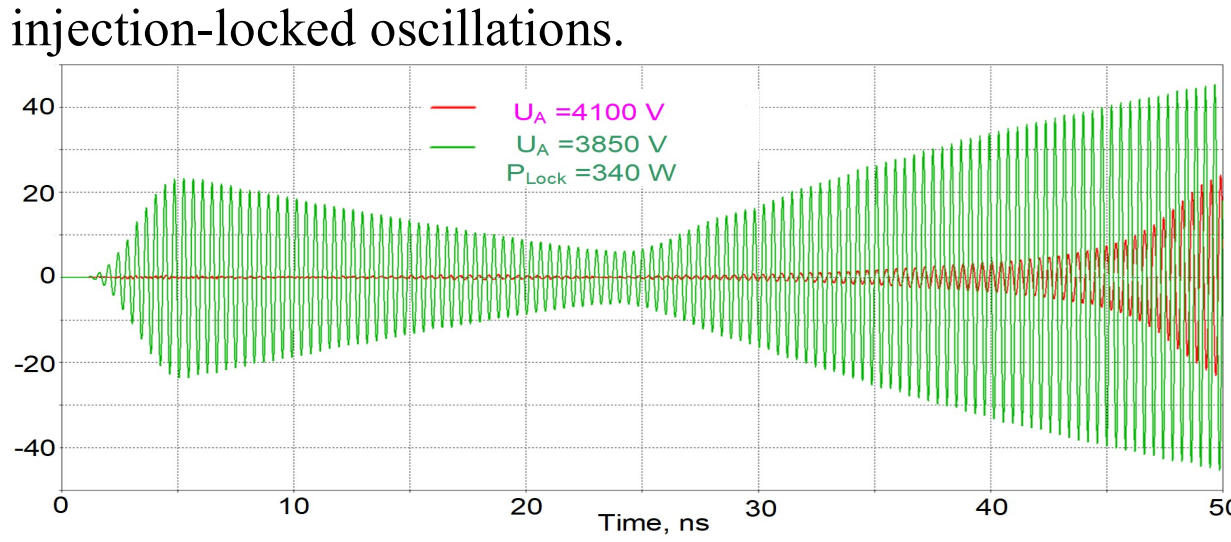

Fig. 8. Red oscillations show buildup of the magnetron free run operation at the anode voltage 4.10 kV. Green oscillations indicate the Landau damping of the synchronizing wave and buildup of the oscillations on output of the magnetron at a forcing signal of about -5 dBc with the anode voltage below the self-excitation

threshold. At such anode voltage the magnetron self-oscillation is impossible.

Figures 7 and 8 show that the delay of phase-locked oscillations in CW magnetrons required for the transient process of the injection locking is sufficiently large and determined simultaneously by two phenomena: Landau damping of the synchronizing wave [7] and a sufficiently slow increase in the RF field in the resonant system caused by insufficient phase grouping of Larmor electrons, until the azimuthal velocities of the drifting charges reach values that provide the required increments of the synchronous wave and the RF field in the resonant system with a Q-factor >>1. A smaller synchronization signal may notably increase the delay of phase-locked oscillations because of these phenomena, reducing the probability of phase locking [15] and limiting the effective bandwidth of magnetron control. Thus, CW magnetrons may provide phase locking (synchronization) with an effective delay during which the magnetron is a source of noise. We developed methods to eliminate this noise (see the following chapters).

## 5. Operation of a CW injection-locked magnetron in the self-excitation mode

We consider operation of a CW magnetron as it is traditionally assumed: in the self-excitation mode, with an injection-locking signal of power $P_{Lock}$.

The effective injection-locking bandwidth $\Delta f$ at locking-signal power $P_{Lock}$ is expressed as shown in [15]:

$$\Delta f = \frac{f_0}{2Q_L}\sqrt{\frac{P_{Lock}}{P_{Mag}}}. \qquad (5)$$

Here: $f_0$ is, in effect, the magnetron average instantaneous frequency; $Q_L$ is the magnetron loaded Q-factor; and $P_{Mag}$ is the magnetron output power. For free-running CW 2.45 GHz magnetrons, the spectral width is about 2–4 MHz at -3 dBc. The spectra consist of numerous startups with instantaneous frequencies filling the spectral line width [16]. Outside the effective bandwidth, the magnetron cannot be injection-locked at the given $P_{Lock}$, as follows from Adler's theory [17].

The probabilities of the injection-locking process $w_{Lock}$ and free-running operation $w_{FR}$ ($P_{Lock} = 0$) for a CW tube as functions of $P_{Lock}$ can then be evaluated as [18]:

$$w_{Lock} \sim \frac{\Delta f}{\Delta f_{FR}}. \qquad (6)$$

$$w_{FR} \sim \frac{\Delta f_{FR} - \Delta f}{\Delta f_{FR}}. \qquad (7)$$

The theoretical estimates of $\Delta f$, $w_{Lock}$ and $w_{FR}$ vs. $P_{Lock}$ for a microwave oven CW 10 cavities tube are shown in Table 2.

Table 2. The theoretical estimates of $\Delta f$, $w_{Lock}$ and $w_{FR}$ vs. $P_{Lock}$.

| $P_{Lock}$ | $\Delta f$, | $w_{Lock}$ | $w_{FR}$ |
|---|---|---|---|
| -10 dB | 3.87 MHz | ~0.97 | ~0.03 |
| -20 dB | 1.22 MHz | ~0.3 | ~0.7 |
| -30 dB | 0.39 MHz | ~0.1 | ~0.9 |

Thus, the probability of injection-locked RF generation of such an RF source at low injection-locking signal may be notably less than the probability of free-running generation caused by noise. If the noise oscillations are incoherent, their magnitude is much smaller than that of the injection-locked (coherent) oscillations.

These incoherent oscillations result in a notable quasi-continuous broadband noise spectrum, which is a disadvantage for operation of self-oscillating magnetrons with a low locking signal in high-Q SRF cavities. Note that if the effective injection-locking bandwidth approaches zero, i.e., the transient time of injection locking tends to infinity [15], then the probability of the injection-locking process also approaches zero.

The impact of injection-locking signal power on the noise was studied by measuring the spectral density of noise power [6] as a function of locking power, $P_{Lock}$, with a 2.45 GHz magnetron of type 2M137-IL at $P_{Lock}$ = 10 W, 30 W, and 100 W, respectively, at a magnetron output power of 1 kW in CW mode. The spectral density of noise power of the generating tube (in the interval 0.1 MHz–1.0 MHz) increases by ≈20 dBc/Hz at $P_{Lock}$ = 10 W (-20 dBc), compared with measurements at $P_{Lock}$ = 100 W (-10 dBc). This indicates a low probability of phase locking at a locking signal of 10 W (-20 dBc) as it was noted above.

## 6. Operation of CW magnetrons above and below the self-excitation threshold voltage

Measured in CW mode at a resolution bandwidth of 5 Hz, the carrier-frequency offsets in Fig. 9 show the carrier-frequency peaks and quasi-continuous noise spectra of the magnetron type 2M137-IL when the tube operates above or below the self-excitation threshold voltage at various injected signals $P_{Lock}$ and output RF powers $P_{Mag}$ [18]..

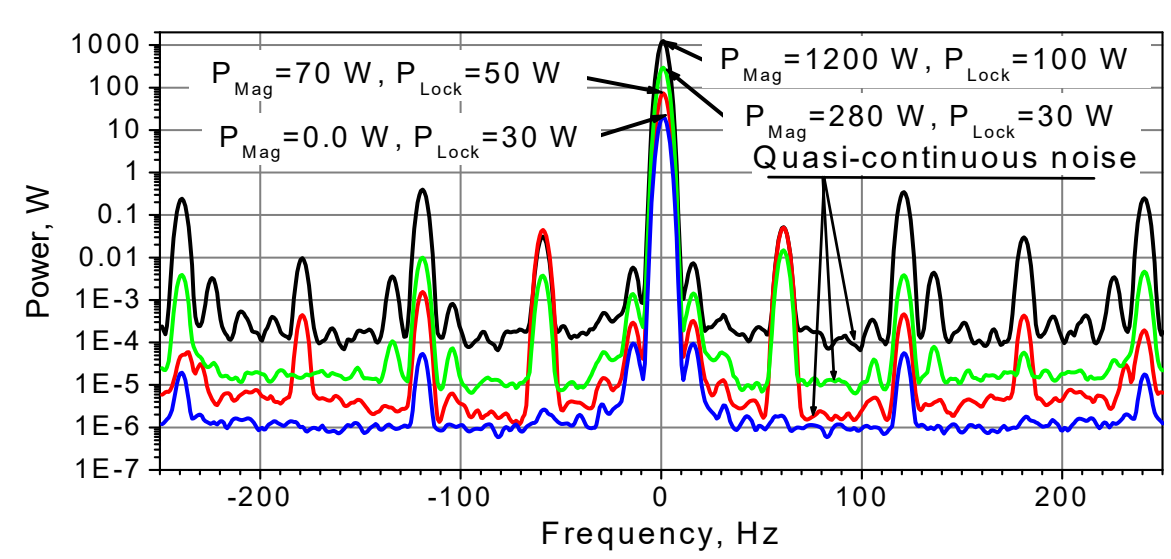

Fig. 9. Offsets of the carrier frequency of the magnetron type 2M137-IL with the threshold of self-excitation of 4.04 kV measured in CW mode at the various output RF powers $P_{Mag}$, and the injection-locking (forcing) signal $P_{Lock}$.

The trace $P_{Mag}$ = 0.0 W, $P_{Lock}$ = 30 W is the frequency offset of the injection-locking signal when the magnetron anode voltage was off. The traces $P_{Mag}$ = 70 W, $P_{Lock}$ = 50 W and $P_{Mag}$ = 280 W, $P_{Lock}$ = 30 W show tube operation below the self-excitation threshold voltage. The trace $P_{Mag}$ = 1200 W, $P_{Lock}$ = 100 W corresponds to operation above the self-excitation voltage. The tube anode voltages $U_{Mag}$ in these measurements were 3.90, 4.01, and 4.09 kV, respectively

Note the very narrow measured spectral line widths of the magnetron, which are highly suitable for powering high-Q SRF cavities.

A noticeable level of quasi-continuous noise of magnetrons operating above the self-excitation threshold voltage indicates a high probability of incoherent generation, i.e., a reduced probability of the injection locking process. For accelerators, this may cause problems with suppressing parasitic modulations of the accelerating field in SRF cavities.

The dependences of the magnetron total noise power and the output power excluding sidebands shown in Fig. 9, as functions of the anode voltage, are plotted in Fig. 10. It is seen that lowering the tube anode voltage reduces the total noise [18]. At $U_{Mag}$ =3.90 kV dominates the regenerative noise.

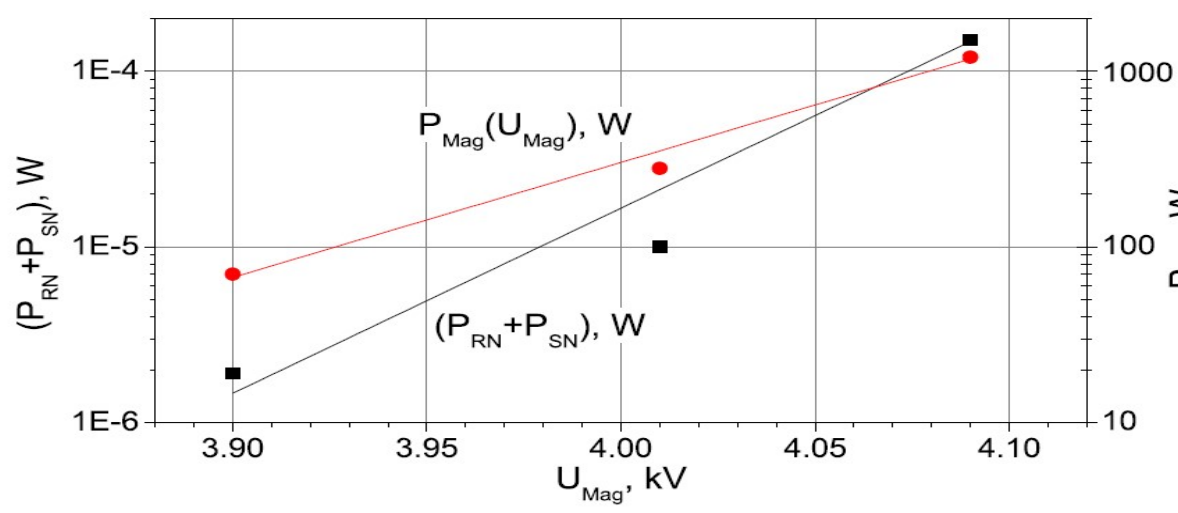

Fig. 10. The dependencies of the magnetron total noise power (black dots, left scale), excluding sidebands, and the output power (red dots, right scale) vs. the anode voltage $U_{Mag}$. RN and SN mean the regenerative and spontaneous noises, respectively. Solid lines are the exponential fits of build-ups of the output power, $P_{Mag}$ and $P_{RN} + P_{SN}$ power for the magnetron 2M137-IL [18].

The ratios of the carrier frequency peak power to the quasi-continuous noise power (excluding sidebands) reach a maximum when the CW magnetron operates below the self-excitation threshold. This is suitable for SRF accelerators.

## 7. Stimulated RF generation mode of a CW magnetron

A developed technique realizing the Stimulated RF generation mode was experimentally verified in a pulse regime using the CW 2.45 GHz magnetron type 2M219G (nominal output power 945 W and measured self-excitation threshold voltage 3.69 kV) [19]. This mode uses a relatively large injected forcing signal, and the anode voltage is chosen to eliminate any startup of the tube by noise or PS ripples, i.e., to eliminate self-oscillations, thereby reducing the regenerative gain. This reduced the tube's regenerative gain magnitude by an order or more. A pulse regime with a train of pulses showing the absence of magnetron startups caused by noise or PS ripples in the intervals between pulse trains was used.

The forcing signal excites a rotating synchronous wave in the interaction space with an azimuthal electric field sufficient for phase locking of the spontaneous oscillations and improves phase grouping of the charges moving in the interaction space. This, in turn, increases the azimuthal electric field of the synchronous wave, and so on. The incoherent spontaneous emission is transformed by phase locking into a coherent oscillation amplified by the resonant system of the magnetron. The setup used to study the Stimulated RF generation mode in a pulse regime with a gated forcing signal was presented in [19]. The main characteristics of RF generation in this mode are presented below.

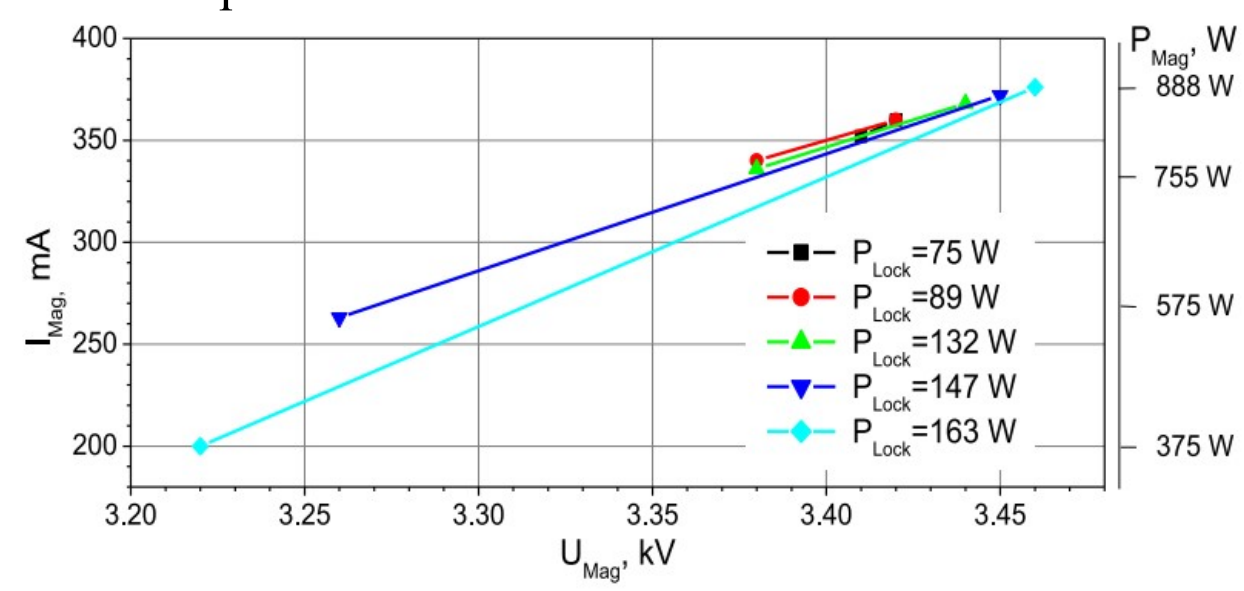

Fig. 11. Measured ranges of the anode voltage and current of the 2M219G magnetron operating in the Stimulated RF generation mode at various power levels of the injected resonant forcing signal $P_{Lock}$. The right scale shows measured RF output power $P_{Mag}$ of the magnetron corresponding to the magnetron current, $I_{Mag}$ [19].

The solid lines in Fig. 11 show the measured ranges of magnetron anode voltage and current as functions of the power of the injected forcing signal $P_{Lock}$. The magnetron was powered by a pulse high-voltage (HV) source [20], using a partial discharge of a 200 µF storage capacitor to provide the pulse anode voltage. The pulse HV source was powered by a charging Glassman 10 kV, 100 mA switching power supply allowing voltage control. The measured ranges of magnetron RF generation were obtained with a forcing RF signal of 2.2 ms duration and a pulse anode voltage of 5.1 ms [19]. The magnetron current was measured by a current transducer of type LA 55-P with a measuring-circuit integration time of about 50 µs.

As follows from Fig. 11, an injected resonant signal of 163 W (-7.6 dBc) provides a magnetron power-control range of ≈3.7 dB by variation of the magnetron current. The increase in the range of tube power (current) control is caused by the larger coherent gain [14] resulting from improved phase grouping at a larger injected resonant forcing signal.

The magnetron performs pulse Stimulated RF generation by injection of a pulsed forcing resonant signal into the tube during DC feeding power, as shown in Figs. 12 and 13. The shapes and values of the forcing and output RF signals were measured by detectors with zero-bias Schottky diodes calibrated to better than ±0.5%.

Despite the significantly lower regenerative gain at a reduced anode voltage, the magnetron in the Stimulated RF generation mode with a large forcing signal converts the spontaneous oscillations in the interaction space into coherent ones [14], which are amplified by the resonant system of the tube. This ensures virtually nominal output RF power with high coherent gain that compensates reduced regenerative gain [ibid.].

The absence of noise in the absence of an injected forcing resonant signal in the Stimulated RF generation mode is represented by the traces in Figs. 12 and 13 [19].

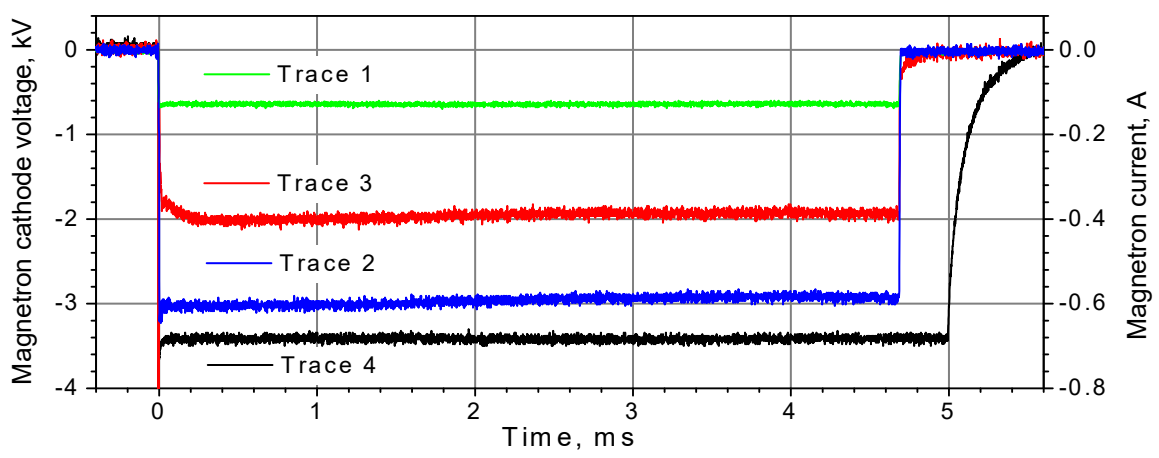

Fig. 12. Pulse operation of the magnetron in the Stimulated RF generation mode with pulse duration of 4.7 ms. Traces 1 and 2 are the resonant injected and the magnetron output RF signals with powers of 74 W and 888 W, respectively; trace 3 is the magnetron pulse current measured by the current transducer (right scale); trace 4 is the magnetron cathode voltage (−3.43 kV).

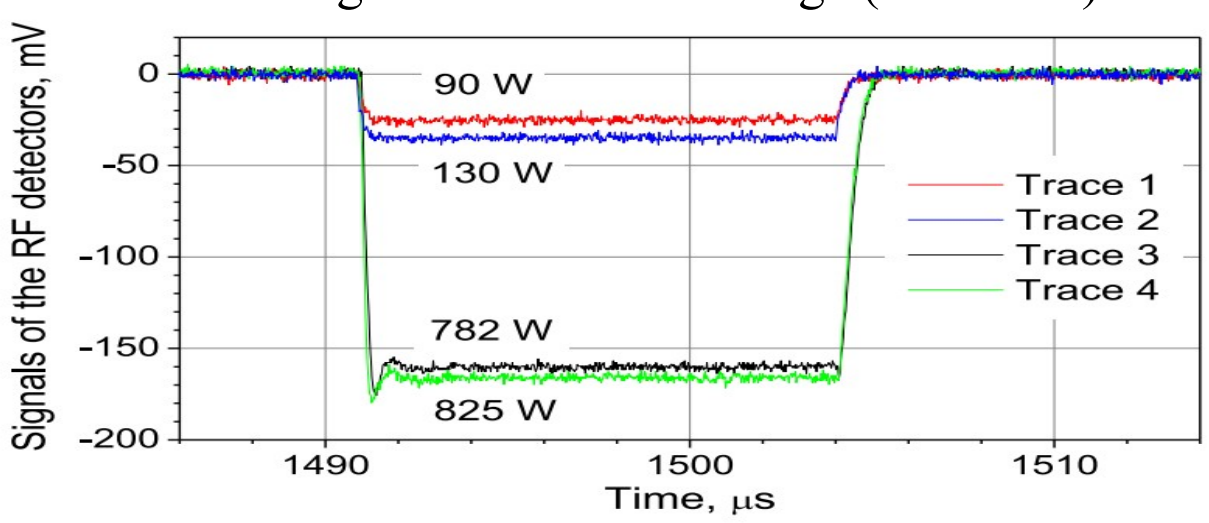

Fig. 13. Measured pulses of the 13 µs 20 kHz train when the magnetron operates in the Stimulated RF generation mode. Traces 3 and 4 are the magnetron output RF signals in dependence on the power of forcing signals, traces 1 and 2, respectively. The measurements were performed at the magnetron anode voltage of 3.41±0.01 kV.

The conversion efficiency $\eta$ defined as the ratio of generated RF power $P_{RF}$ to the consumed power of a CW magnetron operating in the Stimulated RF generation mode, neglecting filament power, is shown below.

$$\eta \approx P_{RF}/(U_{Mag}\cdot I_{Mag} + P_{Lock}) \tag{11}$$

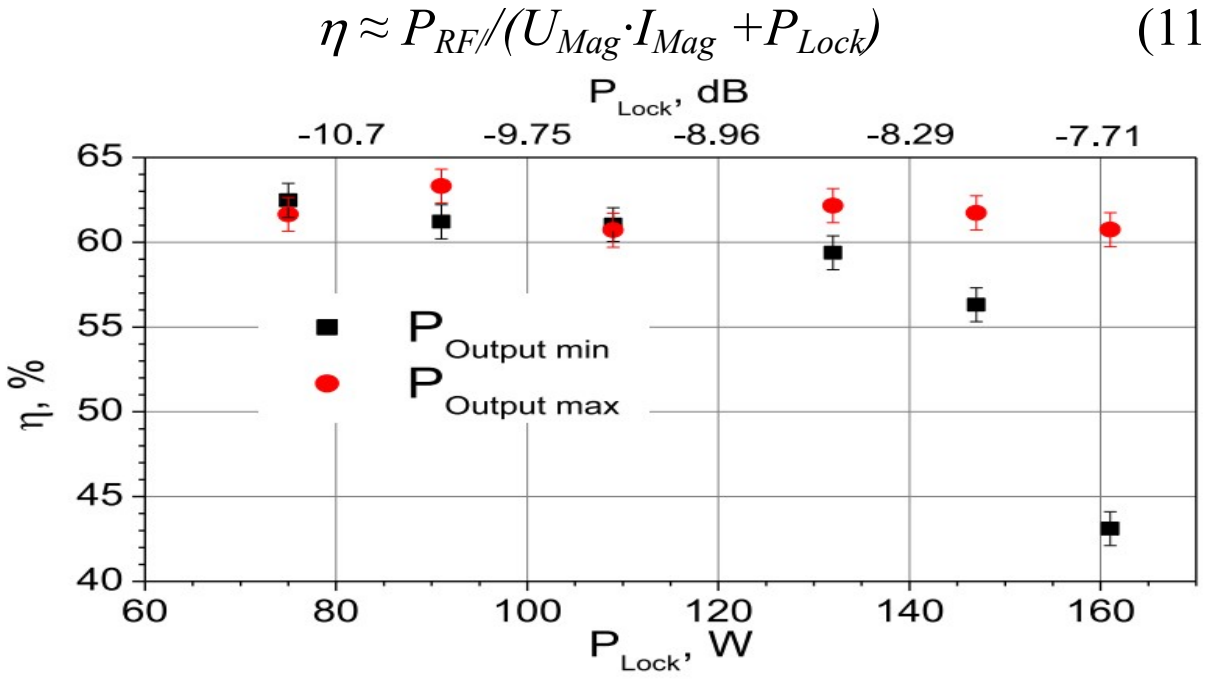

Fig. 14. Dependence of conversion efficiency of the 2M219G magnetron on power of injected forcing signal $P_{Lock}$ [19].

The plotted measured results correspond to the maximum and minimum powers of the magnetron operating in the Stimulated RF generation mode. This demonstrates the range of magnetron power (current) control as a function of $P_{Lock}$. This method of wide-range power control is the most efficient [5]. The measured conversion efficiency of the magnetron operating above the self-excitation threshold in the "free-run" mode ($P_{Lock}$ = 0) at the nominal tube power is ≈54%.

The use of two-cascade magnetrons allows the required power of the forcing signal for the Stimulated RF generation mode to be reduced to ≈ -20 dBc [6]. In this scheme, both magnetrons operate in the Stimulated RF generation mode; the first, with power ≈10% of the required maximum RF power, drives the second high-power tube, while the required power control is realized by varying the current in the high-power magnetron. The cost of two-cascade magnetrons in mass production should be lower than that of traditionally used RF sources.

## 8. Bandwidth of the phase and power control of CW magnetrons in the Stimulated RF generation mode

For various SRF accelerator projects (rings, linacs, ADS facilities, and even industrial SRF accelerators), wideband control of SRF sources in phase and power is important for suppression of various parasitic modulations (microphonics, etc.). The Stimulated RF generation mode uses a sufficiently large resonant forcing signal, which is also necessary for wideband control; it is therefore most suitable for magnetron RF sources for various SRF accelerators.

The control bandwidth $BW_C$ for 2.45 GHz microwave oven magnetrons of type 2M137-IL was obtained by measuring the transfer-function magnitude and phase characteristics [6]. This is presented in Fig. 15 [13].

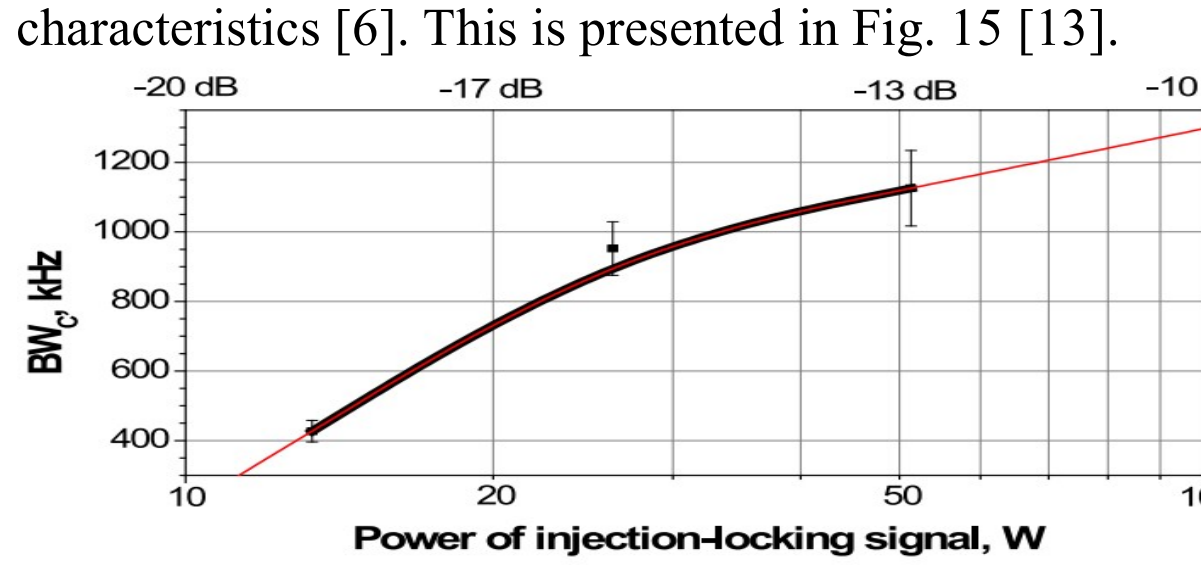

Fig. 15. The admissible bandwidth of control of 2.45 GHz microwave oven magnetrons determined by measured transfer functions characteristics. Black bold line shows the range and results of measurements with B-spline fit; the thin red line shows extrapolation.

As shown by PIC simulations (Fig. 7), the maximum phase-control bandwidth in 10 vans microwave oven magnetrons is limited to a frequency of ~5 MHz (because of the effective phase-synchronization delay), even with a sufficiently high forcing signal. Even higher forcing signal without usage of Stimulated RF generation mode does not provide absolute stabilization of the magnetron.

This means that the bandwidth of CW magnetron control is notably narrower than that of traditionally used RF amplifiers (klystrons, IOTs, solid-state amplifiers). The graph in Fig. 15 shows that the measured control bandwidth for microwave oven magnetron RF sources is more than 1 MHz when they operate in the Stimulated RF generation mode.

The transfer-function magnitude characteristics measured in the phase-modulation domain [21] imply that a low-level RF (LLRF) controller may have a closed loop with a bandwidth of ≥100 kHz and will be able to suppress all expected low-frequency perturbations in the system, such as parasitic low-frequency phase modulation, phase perturbations from SRF-cavity beam loading, cavity dynamic-tuning errors, perturbations caused by magnetron PS ripples, etc.

The impact of high-frequency phase noise of injection-locked magnetrons on the accelerating field in an SRF cavity has been numerically simulated with a model of a proportional-integral (PI) feedback phase loop around a superconducting cavity with broadband noise. In this model, a 200 Hz half-bandwidth low-pass filter represents the cavity baseband response, a 400 kHz bandwidth noise source represents the phase noise of the magnetron, and a 2 μs delay represents the total system group delay. The PI loop is configured with a proportional gain of 200 and an integral gain of $1.2{\cdot}10^7$ rad/s [21].

The numerical modeling demonstrates that broadband noise associated with exaggerated magnetron high-frequency noise is suppressed by the controller with the PI loop, including the SRF cavity, by ≈50 dB in peak-to-peak measurements, Fig. 16. The measurements and simulation indicate that an RF source based on CW magnetrons, injection-locked by a phase-modulated signal and dynamically controlled by a wideband feedback system can provide the phase and amplitude stability of the accelerating field required for SRF cavities in proton/ion accelerators, rings, ADS facilities, and industrial accelerators, as noted in [3].

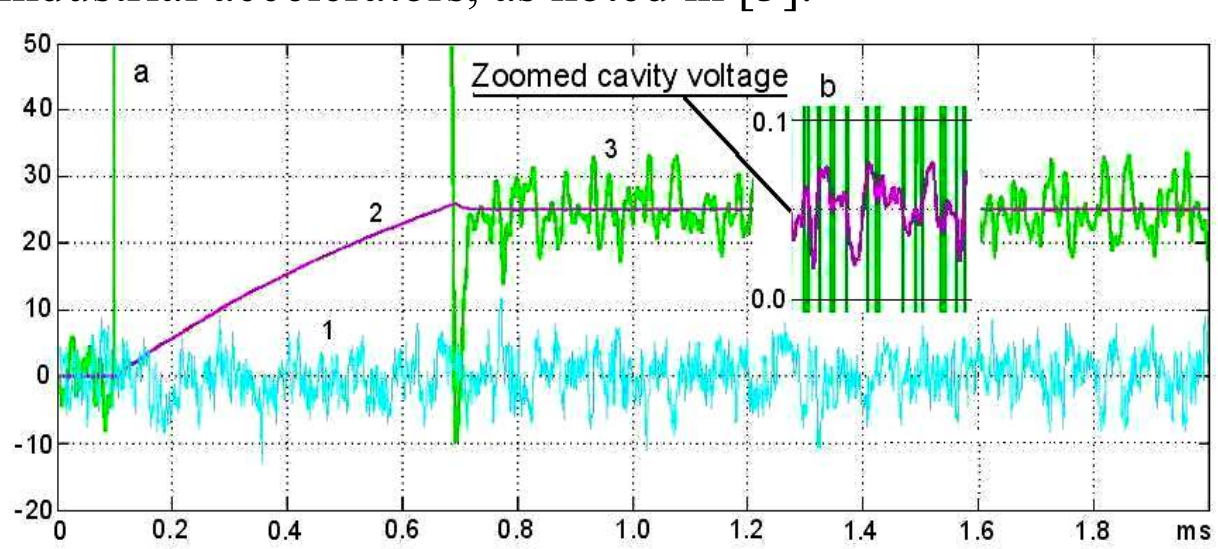

Fig. 16. a: curve 1 shows the bandwidth disturbance of 400 kHz; curve 2 shows cavity voltage; curve 3 shows RF drive. Vertical scale is 10MV/div. Inset "b" shows zoomed in ≈300 times (in vertical) trace of the cavity voltage, curve 2, in time domain. Vertical scale in inset "b" is 0.1MV/div.

Analysis of high-temporal-resolution PIC simulations showing reduce of duration in the transient process of injection-locking at a quite high forcing signal and the characteristics of CW magnetrons operating in the Stimulated RF generation mode reveals that magnetrons capable of suppressing quasi-continuous noise can generate oscillations driven by an injected resonant signal - specifically, forced coherent oscillations analogous to those of conventional RF amplifiers. In other words, the stimulated RF generation mode transforms the regenerative self-oscillation of a CW magnetron into coherent oscillations similar to those produced by conventional master-oscillator-driven RF amplifiers, making them highly suitable for use in various SRF accelerators.

## 9. Summary

The presented study of CW magnetrons for use in SRF accelerators, carried out through analytical and numerical modeling of operation and control and verified by PIC simulations and experiments, demonstrates the applicability of these RF generators for powering of SRF cavities in various accelerators, which increases efficiency, optimizes dynamic control and reduces cost.

The spectral density of noise power of CW microwave oven magnetrons measured at an injected locking signal of -20 dBc shows an increase of ~20 dBc/Hz compared with that measured at a locking signal of -10 dBc. This indicates a low probability of injection-locked RF generation of the magnetron at a locking signal of -20 dBc and a significantly increased probability of the injection-locking process at $P_{Lock} \geq -10$ dBc.

Measurements show that the quasi-continuous noise spectra are significantly reduced in operation of the CW magnetrons below the threshold of self-excitation.

The developed Stimulated RF generation mode eliminates the quasi-continuous noise of CW magnetrons when there is no injected forcing signal; i.e., in this mode the magnetron generates oscillations forced by the injected resonant signal, namely forced coherent oscillations similar to those of traditional RF amplifiers.

The efficiency of CW magnetrons in the Stimulated RF generation mode is higher than that of magnetrons operating above the self-excitation threshold voltage, whether in free-run mode or driven by a small injection-locking signal; see [6].

The Stimulated RF generation mode allows dynamic control of magnetron power with the highest efficiency by changing the magnetron current over a wide range, up to almost the nominal tube power.

The Stimulated RF generation mode is suitable for CW and pulsed SRF accelerators. It provides 100% pulse modulation of the magnetron output power with 100% pulse modulation of the injected forcing signal; thus, HV pulse modulators for pulsed SRF accelerators are not required.

The bandwidth of phase and power control in CW magnetrons operating in the Stimulated RF generation mode is well suited for suppression of parasitic modulations in the SRF cavities of various accelerators.

Numerical modeling based on measured characteristics of the transfer functions of CW magnetrons have demonstrated the applicability of these tubes as RF generators for superconducting accelerators: proton/ion rings, linacs, ADS facilities and industrial accelerators.

Two-cascade magnetrons provide the reduced injected forcing signal required for the Stimulated RF generation mode. This allows the power of the injected resonant signal to be reduced by ≈10 dB. The first CW magnetron, with $P_{Mag}$ ≈10% of the power required from the RF source, provides the injected forcing signal for the second, high-power CW magnetron. Power control is realized by regulating the magnetron current in the high-power tube. In mass production, the cost of the cascaded RF source will increase insignificantly.

The relatively high-power forcing signal used for the Stimulated RF generation mode significantly reduces electron back-streaming that overheats the magnetron cathode; this is important for improving magnetron longevity.

The model we developed for residual gas ionization by Larmor electrons, combined with simulation results regarding cathode sputtering caused by accelerated ions bombardment in powerful CW tubes, indicates that the longevity of CW magnetrons may be significantly extended through the use of built-in ion pumps [18].

## Acknowledgment

This manuscript has been authored by collaboration of Muons, Inc - Fermilab under CRADA-FRA-2017-0026 and CRADA FRA 2023-0029 between Muons, Inc., and Fermilab under FermiForward Discovery Group, LLC, Operator of Fermi National Accelerator Laboratory under contract No. 89243024CSC000002 with the U.S. Department of Energy, Office of Science, Office of High Energy Physics.

**Authored by Jefferson Science Associates, LLC under U.S. DOE Contract No. DE-AC05-06OR23177. The U.S. Government retains a non-exclusive, paid-up, irrevocable, world-wide license to publish or reproduce this manuscript for U.S. Government purposes.